\documentclass[english,aps,manuscript]{article}
\usepackage[T1]{fontenc}
\usepackage[latin9]{inputenc}
\usepackage{geometry}
\usepackage{amstext}
\usepackage{stmaryrd}
\usepackage{graphicx}
\usepackage{setspace}
\makeatletter

\providecommand{\tabularnewline}{\\}

\newcommand{\lyxaddress}[1]{
	\par {\raggedright #1
	\vspace{1.4em}
	\noindent\par}
}

\makeatother

\usepackage{babel}
\begin{document}
\title{Neutrino Mass Eigenvalues for Different Scheme within Four Flavor
Neutrino Framework }
\author{Vivek Kumar Nautiyal$\mathrm{^{1},\,}$Bipin Singh Koranga$\mathrm{^{2},\,}$Jaydip
Singh$^{3}$}
\maketitle

\lyxaddress{$\mathrm{^{1}}$Department of Physics, Babasaheb Bhimrao Ambedkar
University, Lucknow-226025, India}

\lyxaddress{$\mathrm{^{2}}$Department of Physics, Kirori Mal college (University
of Delhi), Delhi-110007, India}

\lyxaddress{$\mathrm{^{3}}$Department of Physics, University of Lucknow, Lucknow-226007,
India}
\begin{abstract}
In this paper, we discuss neutrino mass eigenvalues for four flavor
neutrino mixing. An extra mass states, in four flavor mixing and possible
various combination of CP violating Majorana phases effects the neutrino
mass eigenvalues. We have considered the effective Majorana mass $m_{e}$,
related for $\,\left(\beta\beta\right)_{0\nu\,}$ decay. In calculation,
we consider two different neutrino mass order, normal and inverted.
We find the limits for neutrino mass eigenvalue $m_{i}$ in the different
neutrino mass spectrum and the sum of all four neutrino masses is
$\sum\equiv m_{1}+m_{2}+m_{3}+m_{4}\sim1.17eV\,$ which is relevant
for cosmological observations and explain the different neutrino oscillations
data. 
\end{abstract}

\section{Introduction}

The Dirac and Majorana nature of a particle has been a puzzle for
last decades. Any form of the kinematical test has not been capable
to determine the particle's nature. Our only chance is the experimental
results of Neutrinoless Double Beta Decay via the decay of two-electron
without the emissions of two neutrinos {[}1{]} which eventually confirm
the Majorana nature of the neutrinos. This type of decay violates
the lepton number by 2 units. In order to confirm the type of nature
of a particle it can probe the physics beyond the Standard Model like
the compositeness, violation of equivalnce principle, absolute neutrino
mass scale and the neutrino mass ordering {[}2-5{]}. Only three flavor
of neutrinos are in the standard neutrino oscillation with mass-squared
differences of order $\,10^{-4}\,$ and $\,10^{-3}\,$ eV$^{2}${[}8{]}.
The three neutrino framework has been found to be effective in explaining
numerous sorts of experiments. However, the results of the LSND{[}6{]}
and MiniBooNE{[}7{]} experiments point to the presence of one or two
neutrino states with a considerable mass of eV scale.{[}6, 7{]}. This
additional neutrino termed as \textquotedbl sterile neutrino\textquotedbl .
The effective mass in neutrinoless double beta decay {[}9-11{]} and
the oscillation probabilities in different baseline experiments are
significantly changes due to the presencs of sterile neutrinos. Three
scales of neutrino mass-squared differences $\triangle m_{solar}^{2}\ll\triangle m_{atm}^{2}\ll\triangle m_{LSND}^{2}$
were found by atmospheric and solar neutrino oscillation experiments,
as well as LSND collaboration, which needed four neutrinos with defined
mass to explain these results. 

All these neutrino oscillation data could be explained successfully
by four neutrinos with definite mass by S.M. Bilenky et al. {[}12,
13, 14, 15, 16{]}. The preference of a (2+2) mass spectrum over a
(3+1) spectrum is due to the LSND finding {[}12, 13{]} which was found
to be compatible with null results from accelerator {[}17{]} and reactor
{[}18{]} disappearance. Both atmospheric and solar oscillations might
be influenced by sterile neutrinos. Indeed, The sterile neutrino has
larger contribution in the solar or atmospheric oscillations or in
both for (2+2) mixing schemes while in (3+1) scheme has the three
active conventional neutrinos and one addtional sterile neutrino which
has larger mass than the others. Thus four-neutrino models have been
studied by many authors {[}19, 20, 21, 22, 36, 37, 38{]}.

This paper is organized as follows. In Sec. 2, we briefly describe
the nearly degenerate masses in different scheme for both mass order.
In Sec. 3 we summarize the effective electron neutrino mass in different
scheme for normal and Inverted mass order. Results and Conclusions
are briefly discussed in Sec. 4 and Sec. 5 respectively.

\section{Degenerate masses in Different Scheme}

\begin{figure}
\includegraphics{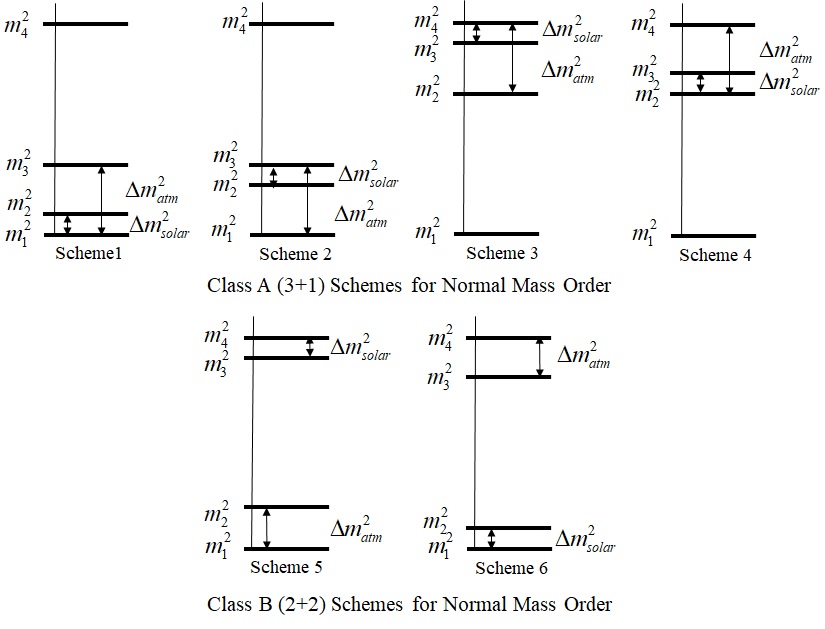}

\caption{Neutrino mass spectrum with Degenerate masses for Normal mass order }
\end{figure}
\begin{figure}
\includegraphics{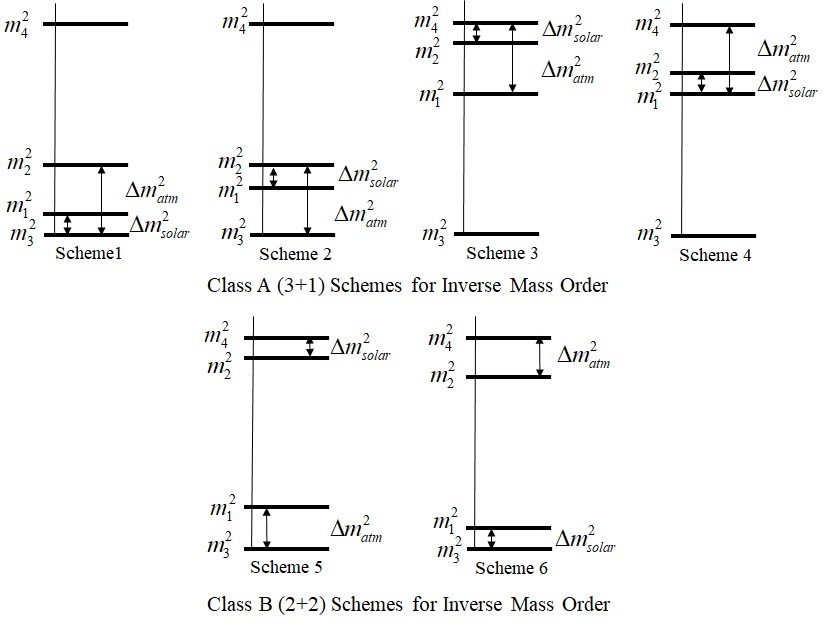}

\caption{Neutrino mass spectrum with Degenerate masses for Inverted mass order}

\end{figure}

The mixing of four neutrinos {[}23-24{]}, which allows to accommodate
all the three existing neutrinos with an extra neutrino $\nu_{s|}$
in four flavour neutrino mixing. Here $\nu_{s|}$ is a sterile neutrino
state. The presence of one sterile neutrino, in four flavour neutrino
mixing provides the differents possibilities for the mass spectrum
of four neutrinos. The neutrino mass-squared differences associated
with solar, atmospheric and LSND for six possibility for normal and
Inverted mass order are shown in Fig. (1) and Fig.(2). These six possible
mass spectra are broadly classified into two classes of schemes. The
class A, which is also called (3+1) scheme, and class B, which is
termed as (2+2) scheme for both mass order are shown in Fig. (1) and
Fig. (2), respectively. The two nearly degenerate pairs of neutrino
mass egienstates $\,\Delta m_{solar}^{2}\,$and$\,\Delta m_{atm}^{2}\,$separated
by the LSND scale $\,\Delta m_{LSND}^{2}=1eV^{2}$are described by
(2+2) scheme. The combination of these two pairs of degenerate masses
$\,\Delta m_{solar}^{2}\,$and$\,\Delta m_{atm}^{2}\,$are described
by scheme 5 and scheme 6 of class B as shown in Fig. (2). The solar
and atmospheric data are explained by the mixing of $\nu_{2},\nu_{1\,}$and
$\nu_{3},\nu_{4\,}$. The additional neutrino in four flavor mixing
has a significant contribution in the solar or atmospheric or in both
the oscillation. (3+1) scheme are assumed the solar and atmospheric
oscillation are given by the three active neutrinos and the addition
neutrino is sterile one which is considered to be heavier than the
three active neutrinos. This addition of sterile neutrino provides
a new independent neutrino mass squared difference,$\,\Delta m_{41\,}^{2}$,
and mixing matrix is elaborated from $\,3\times3\,$ to $\,4\times4\,$
by incorporating three new mixing angles,$\,\theta_{14},\theta_{24},\theta_{34}$~
and two additional CP phases $\,\delta_{14}\,$ and $\,\delta_{24}\,$.
The large mass eigenvalue from the coupling of $\nu_{e}\,$and $\nu_{\mu}\,$via
small mixings with $\nu_{s}\,$, separated from others are capable
for explain the LSND data. Earlier, (3+1) scheme has ruled out due
to its incompatibility with the LSND data and the results of the CDHS
and BUGEY experiments {[}12, 13{]}. However, the shifting of LSND
allowed region from the new LSND experiment {[}2{]} results opens
a small channel for (3+1) scheme. Another one is (1+3) scheme, in
which a massive neutrino is lighter than three neutrinos. This scheme
is disfavoured by the upper bound on the neutrino masses that is smaller
than 1eV, and by the upper bound on effective neutrino mass in neutrinoless
double beta decay if neutrino are Majorana particles {[}25, 26{]}.

The order of mass spectrum can be normal mass order ($m_{4}\gg m_{3}>m_{2}>m_{1}\,\,$and$\,\,\Delta m_{LSND}^{2}=\Delta m_{41}^{2}\,$)
or inverted mass order ($m_{3}<m_{1}<m_{2}\ll m_{4}\,\,$and~$\,\Delta m_{LSND}^{2}=\Delta m_{43}^{2}\,$).
The pairs of degenerate masses $\,\Delta m_{solar}^{2}\,$and$\,\Delta m_{atm}^{2}\,$
with six possible scheme for Normal mass order are describe as {[}12{]};
\begin{equation}
\begin{array}{cccccc}
Class\,A,\\
 & Scheme\,1 & = & \Delta m_{21}^{2}=\Delta m_{solar}^{2} & ; & \Delta m_{31}^{2}=\Delta m_{atm}^{2}\\
 & Scheme\,2 & = & \Delta m_{32}^{2}=\Delta m_{solar}^{2} & ; & \Delta m_{31}^{2}=\Delta m_{atm}^{2}\\
 & Scheme\,3 & = & \Delta m_{43}^{2}=\Delta m_{solar}^{2} & ; & \Delta m_{42}^{2}=\Delta m_{atm}^{2}\\
 & Scheme\,4 & = & \Delta m_{32}^{2}=\Delta m_{solar}^{2} & ; & \Delta m_{42}^{2}=\Delta m_{atm}^{2}\\
Class\,B,\\
 & Scheme\,5 & = & \Delta m_{43}^{2}=\Delta m_{solar}^{2} & ; & \Delta m_{21}^{2}=\Delta m_{atm}^{2}\\
 & Scheme\,6 & = & \Delta m_{21}^{2}=\Delta m_{solar}^{2} & ; & \Delta m_{43}^{2}=\Delta m_{atm}^{2}
\end{array}
\end{equation}

and the pairs of degenerate masses $\,\Delta m_{solar}^{2}\,$and$\,\Delta m_{atm}^{2}\,$
with six possible scheme for Inverted mass order are describe as;
\begin{equation}
\begin{array}{cccccc}
Class\,A,\\
 & Scheme\,1 & = & \Delta m_{13}^{2}=\Delta m_{solar}^{2} & ; & \Delta m_{23}^{2}=\Delta m_{atm}^{2}\\
 & Scheme\,2 & = & \Delta m_{21}^{2}=\Delta m_{solar}^{2} & ; & \Delta m_{23}^{2}=\Delta m_{atm}^{2}\\
 & Scheme\,3 & = & \Delta m_{42}^{2}=\Delta m_{solar}^{2} & ; & \Delta m_{41}^{2}=\Delta m_{atm}^{2}\\
 & Scheme\,4 & = & \Delta m_{21}^{2}=\Delta m_{solar}^{2} & ; & \Delta m_{41}^{2}=\Delta m_{atm}^{2}\\
Class\,B,\\
 & Scheme\,5 & = & \Delta m_{42}^{2}=\Delta m_{solar}^{2} & ; & \Delta m_{13}^{2}=\Delta m_{atm}^{2}\\
 & Scheme\,6 & = & \Delta m_{13}^{2}=\Delta m_{solar}^{2} & ; & \Delta m_{42}^{2}=\Delta m_{atm}^{2}
\end{array}
\end{equation}

Where,
\[
\Delta m_{ij}^{2}=m_{i}^{2}-m_{j}^{2}
\]

\section{Neutrino Mass Eigenvalue in Different Scheme}

In this section, we deals with four flavor framework for (3+1) and
(2+2) schemes by incorporating the sterile neutrino of eV range and
the mixing of this sterile neutrino with three neutrinos is light.
By adding one sterile neutrinos {[}27{]}, there is an increment in
mixing angles and CP violating phases in the PMNS matrix $U_{4\vartimes4}$
which is given by,

\begin{equation}
U=R_{34}R_{24}R_{14}R_{23}R_{13}R_{12}P,
\end{equation}

where the matrices $R_{ij}$are rotations in ij space,

\[
R_{14}=\left(\begin{array}{cccc}
c_{14} & 0 & 0 & s_{14}e^{i\delta}\\
0 & 1 & 0 & 0\\
0 & 0 & 1 & 0\\
-s_{14}e^{-i\delta} & 0 & 0 & c_{14}
\end{array}\right)\,\,\,\,\,\,\,\,\,and\,\,\,\,\,\,\,\,\,R_{34}=\left(\begin{array}{cccc}
1 & 0 & 0 & 0\\
0 & 1 & 0 & 0\\
0 & 0 & c_{34} & s_{34}\\
0 & 0 & -s_{34} & c_{34}
\end{array}\right),
\]

where $s_{ij}=sin\theta_{ij}$,$\,c_{ij}=cos\theta_{ij}$. The diagonal
matrix P contains the three Majorana phases $\alpha,\beta\,\,\,and\,\,\,\gamma:$
\begin{equation}
P=diag(1,e^{i\alpha},e^{i\beta},e^{i\gamma})
\end{equation}

Note that there are in total three Dirac CP-violating phase $\delta_{ij}.$
P is constructed in such a way that only Majorana phases show up in
the effective mass expression. The explicit form of U for (3+1) and
(2+2) schemes are define as;
\begin{equation}
\begin{array}{ccc}
U_{(3+1)}=\left(\begin{array}{cccc}
U_{e1} & U_{e2} & U_{e3} & U_{e4}\\
U_{\mu1} & U_{\mu2} & U_{\mu3} & U_{\mu4}\\
U_{\tau1} & U_{\tau2} & U_{\tau3} & U_{\tau4}\\
U_{s1} & U_{s2} & U_{s3} & U_{s4}
\end{array}\right) & and & U_{(2+2)}=\left(\begin{array}{cccc}
U_{s1} & U_{s2} & U_{s3} & U_{s4}\\
U_{e1} & U_{e2} & U_{e3} & U_{e4}\\
U_{\mu1} & U_{\mu2} & U_{\mu3} & U_{\mu4}\\
U_{\tau1} & U_{\tau2} & U_{\tau3} & U_{\tau4}
\end{array}\right),\end{array}
\end{equation}

In the presence of one sterile neutrino, four flavour neutrino mixing
the electron neutrino is combination of mass eigenstate, $\nu_{i}$
with eigenvalue $m_{i}$

\begin{equation}
\begin{array}{ccc}
\nu_{\alpha}=\sum U_{\alpha j}\nu_{j}\,\,\,j=1,2,3,4 & and & \alpha=e,\mu,\tau,s\end{array}.
\end{equation}

Here $U_{ei}$~are the elements of the $4\vartimes4$ mixing matrix,
which relates the flavour states with the mass eigenvalues. The $\beta\beta_{0\nu}$
decay rate is determined by the effective Majorana mass of the electron
neutrino $m_{e}$. Under the assumption of four flavour neutrino mixing
of neutrino, the effective Majorana neutrino mass $|m_{e}|_{4\nu}$
{[}28{]} is 

\[
|m_{\alpha}|_{4\nu}=|\sum|U_{\alpha j}|^{2}m_{j}|
\]
\begin{equation}
|m_{e}|_{4\nu}=\mid\mid U_{e1}\mid^{2}m_{1}+\mid U_{e2}\mid^{2}e^{i\alpha}m_{2}+\mid U_{e3}\mid^{2}e^{i\beta}m_{3}+\mid U_{e4}\mid^{2}e^{i\gamma}m_{4}|
\end{equation}

Now from Eq.(5), the effective Majorana neutrino mass $|m_{e}|_{4\nu}\,$for
(3+1) scheme is given by;
\begin{equation}
|m_{e}|_{4\nu}=|c_{13}^{2}c_{12}^{2}c_{14}^{2}m_{1}+c_{13}^{2}s_{12}^{2}c_{14}^{2}e^{i\alpha}m_{2}+s_{13}^{2}c_{14}^{2}e^{i\beta}m_{3}+s_{14}^{2}e^{i\gamma}m_{4}|
\end{equation}

and for (2+2) scheme is;
\begin{equation}
\begin{array}{c}
|m_{e}|_{4\nu}=|(-c_{24}c_{23}s_{12}-c_{24}s_{23}s_{13}c_{12}-s_{24}s_{14}c_{13}c_{12})^{2}m_{1}+(c_{24}c_{23}c_{12}-c_{24}s_{23}s_{13}s_{12}\\
-s_{24}s_{14}c_{13}s_{12}e^{i\alpha}m_{2})+(c_{24}s_{23}c_{13}-s_{24}s_{14}s_{13})^{2}e^{i\beta}m_{3}+(s_{24}c_{14})^{2}e^{i\gamma}m_{4}|
\end{array}
\end{equation}

Here, $m_{1},\,m_{2},\,m_{3}\,$and $m_{4}\,$are the masses egienvalues
for vacuum. On squaring Eq. (8), we get
\begin{equation}
\begin{array}{ccc}
|m_{e}|_{4\nu}^{2} & = & \left(|c_{13}^{2}c_{12}^{2}c_{14}^{2}m_{1}+c_{13}^{2}s_{12}^{2}c_{14}^{2}e^{i\alpha}m_{2}+s_{13}^{2}c_{14}^{2}e^{i\beta}m_{3}+s_{14}^{2}e^{i\gamma}m_{4}|\right)\\
 &  & \left(|c_{13}^{2}c_{12}^{2}c_{14}^{2}m_{1}+c_{13}^{2}s_{12}^{2}c_{14}^{2}e^{i\alpha}m_{2}+s_{13}^{2}c_{14}^{2}e^{i\beta}m_{3}+s_{14}^{2}e^{i\gamma}m_{4}|\right)^{*}
\end{array}
\end{equation}
\begin{equation}
\begin{array}{ccc}
|m_{e}|_{4\nu}^{2} & = & \left(c_{13}^{4}c_{12}^{4}c_{14}^{4}m_{1}^{2}+c_{13}^{4}c_{12}^{2}s_{12}^{2}c_{14}^{4}e^{-i\alpha}m_{1}m_{2}+c_{13}^{2}c_{12}^{2}s_{13}^{2}c_{14}^{4}e^{-i\beta}m_{1}m_{3}+c_{13}^{2}c_{12}^{2}c_{14}^{2}s_{14}^{2}e^{-i\gamma}m_{1}m_{4}\right)\\
 &  & \left(c_{13}^{4}c_{12}^{2}s_{12}^{2}c_{14}^{4}e^{i\alpha}m_{1}m_{2}+c_{13}^{4}s_{12}^{4}c_{14}^{4}m_{2}^{2}+c_{13}^{2}s_{13}^{2}s_{12}^{2}c_{14}^{4}e^{i\left(\alpha-\beta\right)}m_{2}m_{3}+c_{13}^{2}s_{12}^{2}c_{14}^{2}s_{14}^{2}e^{i\left(\alpha-\gamma\right)}m_{2}m_{4}\right)\\
 &  & \left(c_{13}^{2}c_{12}^{2}s_{13}^{2}c_{14}^{4}e^{i\beta}m_{1}m_{3}+c_{13}^{2}s_{13}^{2}s_{12}^{2}c_{14}^{4}e^{-i\left(\alpha-\beta\right)}m_{2}m_{3}+s_{13}^{4}c_{14}^{4}m_{3}^{2}+s_{13}^{2}c_{14}^{2}s_{14}^{2}e^{i\left(\beta-\gamma\right)}m_{3}m_{4}\right)\\
 &  & \left(c_{13}^{2}c_{12}^{2}c_{14}^{2}s_{14}^{2}e^{i\gamma}m_{1}m_{4}+c_{13}^{2}s_{12}^{2}c_{14}^{2}s_{14}^{2}e^{-i\left(\alpha-\gamma\right)}m_{2}m_{4}+s_{13}^{2}c_{14}^{2}s_{14}^{2}e^{-i\left(\beta-\gamma\right)}m_{3}m_{4}+s_{14}^{4}m_{4}^{2}\right)
\end{array}
\end{equation}
After some algebra Eq. (11) has the form,

\begin{equation}
\begin{array}{ccc}
|m_{e}|_{4\nu}^{2} & = & A^{2}m_{1}^{2}+B^{2}m_{2}^{2}+C^{2}m_{3}^{2}+D^{2}m_{4}^{2}+2ABm_{1}m_{2}cos\alpha+2ACm_{1}m_{3}cos\beta+2ADm_{1}m_{4}cos\gamma\\
 &  & +2BCm_{2}m_{3}cos\left(\alpha-\beta\right)+2BDm_{2}m_{4}cos\left(\alpha-\gamma\right)+2CDm_{3}m_{4}cos\left(\beta-\gamma\right)
\end{array}
\end{equation}
With,
\begin{equation}
\begin{array}{ccccccc}
A=|U_{e1}|^{2} & , & B=|U_{e2}|^{2} & , & C=|U_{e3}|^{2} & and & D=|U_{e4}|^{2}\end{array}
\end{equation}
The effective Majorana neutrino mass $|m_{e}|_{4\nu}\,$for (2+2)
scheme is same as Equation (12) but the cofficient $\,A,\,B,\,C,\,D\,$are
replaced by the second row of Eq. (5). Now, The effective Majorana
neutrino mass $|m_{e}|_{4\nu}\,$ in term of mass $m_{1}\,$and pairs
of degenerate masses $\,\Delta m_{solar}^{2}\,$and$\,\Delta m_{atm}^{2}\,$define
in Eq. (1) with different scheme for Normal mass order are given by;
\begin{equation}
\begin{array}{cc}
Scheme\,1\\
 & |m_{e}|_{4\nu}^{2}=m_{1}^{2}\left(A^{2}+B^{2}+C^{2}+D^{2}\right)+B^{2}\Delta m_{solar}^{2}+C^{2}\Delta m_{atm}^{2}+D^{2}\Delta m_{LSND}^{2}\\
 & +2ABm_{1}\sqrt{m_{1}^{2}+\Delta m_{solar}^{2}}cos\alpha+2ACm_{1}\sqrt{m_{1}^{2}+\Delta m_{atm}^{2}}cos\beta\\
 & +2ADm_{1}\sqrt{m_{1}^{2}+\Delta m_{LSND}^{2}}cos\gamma+2BC\sqrt{m_{1}^{2}+\Delta m_{solar}^{2}}\sqrt{m_{1}^{2}+\Delta m_{atm}^{2}}cos\left(\alpha-\beta\right)\\
 & +2BD\sqrt{m_{1}^{2}+\Delta m_{solar}^{2}}\sqrt{m_{1}^{2}+\Delta m_{LSND}^{2}}cos\left(\alpha-\gamma\right)\\
 & +2CD\sqrt{m_{1}^{2}+\Delta m_{atm}^{2}}\sqrt{m_{1}^{2}+\Delta m_{LSND}^{2}}cos\left(\beta-\gamma\right)
\end{array}
\end{equation}
\begin{equation}
\begin{array}{cc}
Scheme\,2\\
 & |m_{e}|_{4\nu}^{2}=m_{1}^{2}\left(A^{2}+B^{2}+C^{2}+D^{2}\right)+B^{2}\left(\Delta m_{atm}^{2}-\Delta m_{solar}^{2}\right)+C^{2}\Delta m_{atm}^{2}+D^{2}\Delta m_{LSND}^{2}\\
 & +2ABm_{1}\sqrt{m_{1}^{2}+\Delta m_{atm}^{2}-\Delta m_{solar}^{2}}cos\alpha+2ACm_{1}\sqrt{m_{1}^{2}+\Delta m_{atm}^{2}}cos\beta\\
 & +2ADm_{1}\sqrt{m_{1}^{2}+\Delta m_{LSND}^{2}}cos\gamma+2BC\sqrt{m_{1}^{2}+\Delta m_{atm}^{2}-\Delta m_{solar}^{2}}\sqrt{m_{1}^{2}+\Delta m_{atm}^{2}}cos\left(\alpha-\beta\right)\\
 & +2BD\sqrt{m_{1}^{2}+\Delta m_{atm}^{2}-\Delta m_{solar}^{2}}\sqrt{m_{1}^{2}+\Delta m_{LSND}^{2}}cos\left(\alpha-\gamma\right)\\
 & +2CD\sqrt{m_{1}^{2}+\Delta m_{atm}^{2}}\sqrt{m_{1}^{2}+\Delta m_{LSND}^{2}}cos\left(\beta-\gamma\right)
\end{array}
\end{equation}
\begin{equation}
\begin{array}{cc}
Scheme\,3\\
 & |m_{e}|_{4\nu}^{2}=m_{1}^{2}\left(A^{2}+B^{2}+C^{2}+D^{2}\right)+B^{2}\left(\Delta m_{LSND}^{2}-\Delta m_{atm}^{2}\right)+C^{2}\left(\Delta m_{LSND}^{2}-\Delta m_{solar}^{2}\right)\\
 & +D^{2}\Delta m_{LSND}^{2}+2ABm_{1}\sqrt{m_{1}^{2}+\Delta m_{LSND}^{2}-\Delta m_{atm}^{2}}cos\alpha\\
 & +2ACm_{1}\sqrt{m_{1}^{2}+\Delta m_{LSND}^{2}-\Delta m_{solar}^{2}}cos\beta+2ADm_{1}\sqrt{m_{1}^{2}+\Delta m_{LSND}^{2}}cos\gamma\\
 & +2BC\sqrt{m_{1}^{2}+\Delta m_{LSND}^{2}-\Delta m_{atm}^{2}}\sqrt{m_{1}^{2}+\Delta m_{LSND}^{2}-\Delta m_{solar}^{2}}cos\left(\alpha-\beta\right)\\
 & +2BD\sqrt{m_{1}^{2}+\Delta m_{LSND}^{2}-\Delta m_{atm}^{2}}\sqrt{m_{1}^{2}+\Delta m_{LSND}^{2}}cos\left(\alpha-\gamma\right)\\
 & +2CD\sqrt{m_{1}^{2}+\Delta m_{LSND}^{2}-\Delta m_{solar}^{2}}\sqrt{m_{1}^{2}+\Delta m_{LSND}^{2}}cos\left(\beta-\gamma\right)
\end{array}
\end{equation}
\begin{equation}
\begin{array}{cc}
Scheme\,4\\
 & |m_{e}|_{4\nu}^{2}=m_{1}^{2}\left(A^{2}+B^{2}+C^{2}+D^{2}\right)+B^{2}\left(\Delta m_{LSND}^{2}-\Delta m_{atm}^{2}\right)+C^{2}\left(\Delta m_{LSND}^{2}-\Delta m_{atm}^{2}+\Delta m_{solar}^{2}\right)\\
 & +D^{2}\Delta m_{LSND}^{2}+2ABm_{1}\sqrt{m_{1}^{2}+\Delta m_{LSND}^{2}-\Delta m_{atm}^{2}}cos\alpha\\
 & +2ACm_{1}\sqrt{m_{1}^{2}+\Delta m_{LSND}^{2}-\Delta m_{atm}^{2}+\Delta m_{solar}^{2}}cos\beta+2ADm_{1}\sqrt{m_{1}^{2}+\Delta m_{LSND}^{2}}cos\gamma\\
 & +2BC\sqrt{m_{1}^{2}+\Delta m_{LSND}^{2}-\Delta m_{atm}^{2}}\sqrt{m_{1}^{2}+\Delta m_{LSND}^{2}-\Delta m_{atm}^{2}+\Delta m_{solar}^{2}}cos\left(\alpha-\beta\right)\\
 & +2BD\sqrt{m_{1}^{2}+\Delta m_{LSND}^{2}-\Delta m_{atm}^{2}}\sqrt{m_{1}^{2}+\Delta m_{LSND}^{2}}cos\left(\alpha-\gamma\right)\\
 & +2CD\sqrt{m_{1}^{2}+\Delta m_{LSND}^{2}-\Delta m_{atm}^{2}+\Delta m_{solar}^{2}}\sqrt{m_{1}^{2}+\Delta m_{LSND}^{2}}cos\left(\beta-\gamma\right)
\end{array}
\end{equation}
\begin{equation}
\begin{array}{cc}
Scheme\,5\\
 & |m_{e}|_{4\nu}^{2}=m_{1}^{2}\left(A^{2}+B^{2}+C^{2}+D^{2}\right)+B^{2}\Delta m_{atm}^{2}+C^{2}\left(\Delta m_{LSND}^{2}-\Delta m_{solar}^{2}\right)+D^{2}\Delta m_{LSND}^{2}\\
 & +2ABm_{1}\sqrt{m_{1}^{2}+\Delta m_{atm}^{2}}cos\alpha+2ACm_{1}\sqrt{m_{1}^{2}+\Delta m_{LSND}^{2}-\Delta m_{solar}^{2}}cos\beta\\
 & +2ADm_{1}\sqrt{m_{1}^{2}+\Delta m_{LSND}^{2}}cos\gamma+2BC\sqrt{m_{1}^{2}+\Delta m_{atm}^{2}}\sqrt{m_{1}^{2}+\Delta m_{LSND}^{2}-\Delta m_{solar}^{2}}cos\left(\alpha-\beta\right)\\
 & +2BD\sqrt{m_{1}^{2}+\Delta m_{atm}^{2}}\sqrt{m_{1}^{2}+\Delta m_{LSND}^{2}}cos\left(\alpha-\gamma\right)\\
 & +2CD\sqrt{m_{1}^{2}+\Delta m_{LSND}^{2}-\Delta m_{solar}^{2}}\sqrt{m_{1}^{2}+\Delta m_{LSND}^{2}}cos\left(\beta-\gamma\right)
\end{array}
\end{equation}
\begin{equation}
\begin{array}{cc}
Scheme\,6\\
 & |m_{e}|_{4\nu}^{2}=m_{1}^{2}\left(A^{2}+B^{2}+C^{2}+D^{2}\right)+B^{2}\Delta m_{solar}^{2}+C^{2}\left(\Delta m_{LSND}^{2}-\Delta m_{atm}^{2}\right)+D^{2}\Delta m_{LSND}^{2}\\
 & +2ABm_{1}\sqrt{m_{1}^{2}+\Delta m_{solar}^{2}}cos\alpha+2ACm_{1}\sqrt{m_{1}^{2}+\Delta m_{LSND}^{2}-\Delta m_{atm}^{2}}ccos\beta\\
 & +2ADm_{1}\sqrt{m_{1}^{2}+\Delta m_{LSND}^{2}}cos\gamma+2BC\sqrt{m_{1}^{2}+\Delta m_{solar}^{2}}\sqrt{m_{1}^{2}+\Delta m_{LSND}^{2}-\Delta m_{atm}^{2}}cos\left(\alpha-\beta\right)\\
 & +2BD\sqrt{m_{1}^{2}+\Delta m_{solar}^{2}}\sqrt{m_{1}^{2}+\Delta m_{LSND}^{2}}cos\left(\alpha-\gamma\right)\\
 & +2CD\sqrt{m_{1}^{2}+\Delta m_{LSND}^{2}-\Delta m_{atm}^{2}}\sqrt{m_{1}^{2}+\Delta m_{LSND}^{2}}cos\left(\beta-\gamma\right)
\end{array}
\end{equation}

For scheme 5 and scheme 6 the coefficient $\,A,\,B,\,C,\,D\,$are
replaced by the second row of Eq. (5). 

Similarly, The effective Majorana neutrino mass $|m_{e}|_{4\nu}\,$
in term of mass $m_{3}\,$and pairs of degenerate masses $\,\Delta m_{solar}^{2}\,$and$\,\Delta m_{atm}^{2}\,$define
in Eq. (2) with different scheme for Inverted mass order are given
by;

\begin{equation}
\begin{array}{cc}
Scheme\,1\\
 & |m_{e}|_{4\nu}^{2}=m_{3}^{2}\left(A^{2}+B^{2}+C^{2}+D^{2}\right)+A^{2}\Delta m_{solar}^{2}+B^{2}\Delta m_{atm}^{2}+D^{2}\Delta m_{LSND}^{2}\\
 & +2AB\sqrt{m_{3}^{2}+\Delta m_{solar}^{2}}\sqrt{m_{3}^{2}+\Delta m_{atm}^{2}}cos\alpha+2ACm_{3}\sqrt{m_{3}^{2}+\Delta m_{solar}^{2}}cos\beta\\
 & +2AD\sqrt{m_{3}^{2}+\Delta m_{solar}^{2}}\sqrt{m_{3}^{2}+\Delta m_{LSND}^{2}}cos\gamma+2BCm_{3}\sqrt{m_{3}^{2}+\Delta m_{atm}^{2}}cos\left(\alpha-\beta\right)\\
 & +2BD\sqrt{m_{3}^{2}+\Delta m_{atm}^{2}}\sqrt{m_{3}^{2}+\Delta m_{LSND}^{2}}cos\left(\alpha-\gamma\right)\\
 & +2CDm_{3}\sqrt{m_{3}^{2}+\Delta m_{LSND}^{2}}cos\left(\beta-\gamma\right)
\end{array}
\end{equation}
\begin{equation}
\begin{array}{cc}
Scheme\,2\\
 & |m_{e}|_{4\nu}^{2}=m_{3}^{2}\left(A^{2}+B^{2}+C^{2}+D^{2}\right)+A^{2}\left(\Delta m_{atm}^{2}-\Delta m_{solar}^{2}\right)+B^{2}\Delta m_{atm}^{2}+D^{2}\Delta m_{LSND}^{2}\\
 & +2AB\sqrt{m_{3}^{2}+\Delta m_{atm}^{2}-\Delta m_{solar}^{2}}\sqrt{m_{3}^{2}+\Delta m_{atm}^{2}}cos\alpha+2ACm_{3}\sqrt{m_{3}^{2}+\Delta m_{atm}^{2}-\Delta m_{solar}^{2}}cos\beta\\
 & +2AD\sqrt{m_{3}^{2}+\Delta m_{atm}^{2}-\Delta m_{solar}^{2}}\sqrt{m_{3}^{2}+\Delta m_{LSND}^{2}}cos\gamma+2BCm_{3}\sqrt{m_{3}^{2}+\Delta m_{atm}^{2}}cos\left(\alpha-\beta\right)\\
 & +2BD\sqrt{m_{3}^{2}+\Delta m_{atm}^{2}}\sqrt{m_{3}^{2}+\Delta m_{LSND}^{2}}cos\left(\alpha-\gamma\right)\\
 & +2CDm_{3}\sqrt{m_{3}^{2}+\Delta m_{LSND}^{2}}cos\left(\beta-\gamma\right)
\end{array}
\end{equation}
\begin{equation}
\begin{array}{cc}
Scheme\,3\\
 & |m_{e}|_{4\nu}^{2}=m_{3}^{2}\left(A^{2}+B^{2}+C^{2}+D^{2}\right)+A^{2}\left(m_{LSND}^{2}-\Delta m_{atm}^{2}\right)+B^{2}\left(m_{LSND}^{2}-\Delta m_{solar}^{2}\right)\\
 & +D^{2}\Delta m_{LSND}^{2}+2AB\sqrt{m_{3}^{2}+m_{LSND}^{2}-\Delta m_{atm}^{2}}\sqrt{m_{3}^{2}+m_{LSND}^{2}-\Delta m_{solar}^{2}}cos\alpha\\
 & +2ACm_{3}\sqrt{m_{3}^{2}+m_{LSND}^{2}-\Delta m_{atm}^{2}}cos\beta+2BCm_{3}\sqrt{m_{3}^{2}+m_{LSND}^{2}-\Delta m_{solar}^{2}}cos\left(\alpha-\beta\right)\\
 & +2AD\sqrt{m_{3}^{2}+m_{LSND}^{2}-\Delta m_{atm}^{2}}\sqrt{m_{3}^{2}+\Delta m_{LSND}^{2}}cos\gamma\\
 & +2BD\sqrt{m_{3}^{2}+m_{LSND}^{2}-\Delta m_{solar}^{2}}\sqrt{m_{3}^{2}+\Delta m_{LSND}^{2}}cos\left(\alpha-\gamma\right)\\
 & +2CDm_{3}\sqrt{m_{3}^{2}+\Delta m_{LSND}^{2}}cos\left(\beta-\gamma\right)
\end{array}
\end{equation}
\begin{equation}
\begin{array}{cc}
Scheme\,4\\
 & |m_{e}|_{4\nu}^{2}=m_{3}^{2}\left(A^{2}+B^{2}+C^{2}+D^{2}\right)+A^{2}\left(m_{LSND}^{2}-\Delta m_{atm}^{2}\right)+B^{2}\left(\Delta m_{LSND}^{2}-\Delta m_{atm}^{2}+\Delta m_{solar}^{2}\right)\\
 & +D^{2}\Delta m_{LSND}^{2}+2AB\sqrt{m_{3}^{2}+m_{LSND}^{2}-\Delta m_{atm}^{2}}\sqrt{m_{3}^{2}+\Delta m_{LSND}^{2}-\Delta m_{atm}^{2}+\Delta m_{solar}^{2}}cos\alpha\\
 & +2ACm_{3}\sqrt{m_{3}^{2}+m_{LSND}^{2}-\Delta m_{atm}^{2}}cos\beta+2AD\sqrt{m_{3}^{2}+m_{LSND}^{2}-\Delta m_{atm}^{2}}\sqrt{m_{3}^{2}+\Delta m_{LSND}^{2}}cos\gamma\\
 & +2BCm_{3}\sqrt{m_{3}^{2}+\Delta m_{LSND}^{2}-\Delta m_{atm}^{2}+\Delta m_{solar}^{2}}cos\left(\alpha-\beta\right)\\
 & +2BD\sqrt{m_{3}^{2}+\Delta m_{LSND}^{2}-\Delta m_{atm}^{2}+\Delta m_{solar}^{2}}\sqrt{m_{3}^{2}+\Delta m_{LSND}^{2}}cos\left(\alpha-\gamma\right)\\
 & +2CDm_{3}\sqrt{m_{3}^{2}+\Delta m_{LSND}^{2}}cos\left(\beta-\gamma\right)
\end{array}
\end{equation}
\begin{equation}
\begin{array}{cc}
Scheme\,5\\
 & |m_{e}|_{4\nu}^{2}=m_{3}^{2}\left(A^{2}+B^{2}+C^{2}+D^{2}\right)+A^{2}\Delta m_{atm}^{2}+B^{2}\left(m_{LSND}^{2}-\Delta m_{solar}^{2}\right)\\
 & +D^{2}\Delta m_{LSND}^{2}+2AB\sqrt{m_{3}^{2}+\Delta m_{atm}^{2}}\sqrt{m_{3}^{2}+m_{LSND}^{2}-\Delta m_{solar}^{2}}cos\alpha\\
 & +2ACm_{3}\sqrt{m_{3}^{2}+\Delta m_{atm}^{2}}cos\beta+2BCm_{3}\sqrt{m_{3}^{2}+m_{LSND}^{2}-\Delta m_{solar}^{2}}cos\left(\alpha-\beta\right)\\
 & +2AD\sqrt{m_{3}^{2}+\Delta m_{atm}^{2}}\sqrt{m_{3}^{2}+\Delta m_{LSND}^{2}}cos\gamma+2CDm_{3}\sqrt{m_{3}^{2}+\Delta m_{LSND}^{2}}cos\left(\beta-\gamma\right)\\
 & +2BD\sqrt{m_{3}^{2}+m_{LSND}^{2}-\Delta m_{solar}^{2}}\sqrt{m_{3}^{2}+\Delta m_{LSND}^{2}}cos\left(\alpha-\gamma\right)
\end{array}
\end{equation}
\begin{equation}
\begin{array}{cc}
Scheme\,6\\
 & |m_{e}|_{4\nu}^{2}=m_{3}^{2}\left(A^{2}+B^{2}+C^{2}+D^{2}\right)+A^{2}\Delta m_{solar}^{2}+B^{2}\left(m_{LSND}^{2}-\Delta m_{atm}^{2}\right)\\
 & +D^{2}\Delta m_{LSND}^{2}+2AB\sqrt{m_{3}^{2}+\Delta m_{solar}^{2}}\sqrt{m_{3}^{2}+m_{LSND}^{2}-\Delta m_{atm}^{2}}cos\alpha\\
 & +2ACm_{3}\sqrt{m_{3}^{2}+\Delta m_{solar}^{2}}cos\beta+2BCm_{3}\sqrt{m_{3}^{2}+m_{LSND}^{2}-\Delta m_{atm}^{2}}cos\left(\alpha-\beta\right)\\
 & +2AD\sqrt{m_{3}^{2}+\Delta m_{solar}^{2}}\sqrt{m_{3}^{2}+\Delta m_{LSND}^{2}}cos\gamma+2CDm_{3}\sqrt{m_{3}^{2}+\Delta m_{LSND}^{2}}cos\left(\beta-\gamma\right)\\
 & +2BD\sqrt{m_{3}^{2}+m_{LSND}^{2}-\Delta m_{atm}^{2}}\sqrt{m_{3}^{2}+\Delta m_{LSND}^{2}}cos\left(\alpha-\gamma\right)
\end{array}
\end{equation}
Again, for scheme 5 and scheme 6 the coefficient $\,A,\,B,\,C,\,D\,$are
replaced by the second row of Eq. (5).

\section{Numerical Results}

In this section, We have calculated the mass eigenvalues  in four
flavor neutrino oscillation for both mass order. The effective majorana
neutrino mass $\,m_{e\,}$ is directly propotional to the inverse
of the half-life of nucleus participated in the process of neutrinless
double beta decay. The effective Majorana mass is not very well understood
but lower and upper bound of the effective majorana neutrino mass
for normal and inverted order are$\,\,0.0089\,eV\leq m_{e}\leq0.0126\,eV\,${[}29{]}
and~$\,0.02\,eV\leq m_{e}\leq0.05\,eV\,$ {[}30{]}, respectively.
The effective neutrino mass eqaution depends on the values of the
Majorana phase $\alpha,$ $\beta,$$\gamma$. We have taken $\Delta m_{31}^{2}=0.002eV^{2}\,${[}30,
31{]},~ $\Delta m_{21}^{2}=0.00008eV^{2}\,${[}33{]} and $\Delta m_{41}^{2}=1.18eV^{2}\,[30]$.
The active sterile neutrino mixing angle are $\,\theta_{14},\,\,\theta_{24}\,\,and\,\,\theta_{34}$.
In this work, we consider following value for sterile neutrino mixing
angles {[}35{]}, $\theta_{14}=3.6^{o},\,\,\,\,\,\,\,\,\,\,\theta_{24}=4^{o},\,\,\,\,\,\,\,\,\,\,\theta_{34}=18.5^{o}.\,$The
constraint for the cosmological observations of neutrino mass, we
have taken sum of all the four masses $\text{\,}\text{\ensuremath{\sum\equiv}}m_{1}+m_{2}+m_{3}+m_{4}.\,$

\begin{table}
\caption{Neutrino mass eigenvalues in eV for Normal mass order.}

\begin{tabular}{lllllllllll}
 &  &  &  &  &  &  &  &  &  & \tabularnewline
\hline 
Majorana &  & Effective mass & Mass &  &  &  & Scheme &  &  & \tabularnewline
\cline{6-11} \cline{7-11} \cline{8-11} \cline{9-11} \cline{10-11} \cline{11-11} 
Phases &  & ($m_{e}$ in eV) & states &  & 1 & 2 & 3 & 4 & 5 & 6\tabularnewline
\hline 
 &  &  &  &  &  &  &  &  &  & \tabularnewline
$\alpha=0^{0},$ &  & 0.0095 & $m_{1}$ &  & 0.01642 & 0.01697 & 0.07235 & 0.07201 & 0.30708 & 0.30708\tabularnewline
$\beta=0^{0},$ &  &  & $m_{2}$ &  & 0.01965 & 0.02034 & 0.07300 & 0.07262 & 0.30794 & 0.30794\tabularnewline
$\gamma=0^{0}$ &  &  & $m_{3}$ &  & 0.04887 & 0.04948 & 0.08636 & 0.08560 & 0.31104 & 0.31104\tabularnewline
 &  &  & $m_{4}$ &  & 1.0865 & 1.0865 & 1.0890 & 1.0888 & 1.1290 & 1.1290\tabularnewline
 &  & 0.011 & $m_{1}$ &  & 0.01782 & 0.01833 & 0.07251 & 0.07217 & 0.30759 & 0.30759\tabularnewline
 &  &  & $m_{2}$ &  & 0.02069 & 0.02192 & 0.07318 & 0.07279 & 0.30772 & 0.30772\tabularnewline
 &  &  & $m_{3}$ &  & 0.04934 & 0.05041 & 0.08671 & 0.08594 & 0.31082 & 0.31082\tabularnewline
 &  &  & $m_{4}$ &  & 1.0865 & 1.0865 & 1.0891 & 1.0889 & 1.1289 & 1.1289\tabularnewline
 &  & 0.0125 & $m_{1}$ &  & 0.02050 & 0.02036 & 0.07269 & 0.07235 & 0.30734 & 0.30734\tabularnewline
 &  &  & $m_{2}$ &  & 0.02299 & 0.02374 & 0.07338 & 0.07299 & 0.30747 & 0.30747\tabularnewline
 &  &  & $m_{3}$ &  & 0.05049 & 0.05146 & 0.08712 & 0.08633 & 0.31058 & 0.31058\tabularnewline
 &  &  & $m_{4}$ &  & 1.0865 & 1.0866 & 1.0891 & 1.0890 & 1.1289 & 1.1289\tabularnewline
 &  &  &  &  &  &  &  &  &  & \tabularnewline
$\alpha=0^{0},$ &  & 0.0095 & $m_{1}$ &  & 0.01642 & 0.01697 & 0.07235 & 0.07201 & 0.30708 & 0.30708\tabularnewline
$\beta=0^{0}$ &  &  & $m_{2}$ &  & 0.01965 & 0.02034 & 0.07300 & 0.07262 & 0.30794 & 0.30794\tabularnewline
$\gamma=180^{0}$ &  &  & $m_{3}$ &  & 0.04887 & 0.04948 & 0.08636 & 0.08560 & 0.31104 & 0.31104\tabularnewline
 &  &  & $m_{4}$ &  & 1.0865 & 1.0865 & 1.0890 & 1.0888 & 1.1290 & 1.1290\tabularnewline
 &  & 0.011 & $m_{1}$ &  & 0.01918 & 0.01833 & 0.07251 & 0.07217 & 0.30759 & 0.30759\tabularnewline
 &  &  & $m_{2}$ &  & 0.02193 & 0.02192 & 0.07318 & 0.07279 & 0.30772 & 0.30772\tabularnewline
 &  &  & $m_{3}$ &  & 0.04997 & 0.05041 & 0.08671 & 0.08594 & 0.31082 & 0.31082\tabularnewline
 &  &  & $m_{4}$ &  & 1.0865 & 1.0865 & 1.0891 & 1.0889 & 1.1289 & 1.1289\tabularnewline
 &  & 0.0125 & $m_{1}$ &  & 0.02050 & 0.02036 & 0.07269 & 0.07235 & 0.30734 & 0.30734\tabularnewline
 &  &  & $m_{2}$ &  & 0.02299 & 0.02374 & 0.07338 & 0.07299 & 0.30747 & 0.30747\tabularnewline
 &  &  & $m_{3}$ &  & 0.05049 & 0.05146 & 0.08712 & 0.08633 & 0.31058 & 0.31058\tabularnewline
 &  &  & $m_{4}$ &  & 1.0865 & 1.0866 & 1.0891 & 1.0890 & 1.1289 & 1.1289\tabularnewline
 &  &  &  &  &  &  &  &  &  & \tabularnewline
$\alpha=0^{0},$ &  & 0.0095 & $m_{1}$ &  & 0.01814 & 0.01850 & 0.07235 & 0.07201 & 0.58099 & 0.58099\tabularnewline
$\beta=180^{0}$ &  &  & $m_{2}$ &  & 0.02124 & 0.02171 & 0.07300 & 0.07262 & 0.58108 & 0.58108\tabularnewline
$\gamma=0^{0}$ &  &  & $m_{3}$ &  & 0.04974 & 0.05018 & 0.08636 & 0.08560 & 0.58322 & 0.58322\tabularnewline
 &  &  & $m_{4}$ &  & 1.0865 & 1.0865 & 1.0890 & 1.0888 & 1.2595 & 1.2595\tabularnewline
 &  & 0.011 & $m_{1}$ &  & 0.02100 & 0.02001 & 0.07251 & 0.07217 & 0.58301 & 0.58301\tabularnewline
 &  &  & $m_{2}$ &  & 0.02363 & 0.02342 & 0.07318 & 0.07279 & 0.58309 & 0.58309\tabularnewline
 &  &  & $m_{3}$ &  & 0.05097 & 0.05120 & 0.08671 & 0.08594 & 0.58524 & 0.58524\tabularnewline
 &  &  & $m_{4}$ &  & 1.0866 & 1.0865 & 1.0891 & 1.0889 & 1.2617 & 1.2617\tabularnewline
 &  & 0.0125 & $m_{1}$ &  & 0.02194 & 0.02214 & 0.07269 & 0.07235 & 0.58531 & 0.58531\tabularnewline
 &  &  & $m_{2}$ &  & 0.02432 & 0.02535 & 0.07338 & 0.07299 & 0.58540 & 0.58540\tabularnewline
 &  &  & $m_{3}$ &  & 0.05122 & 0.05236 & 0.08712 & 0.08633 & 0.58755 & 0.58755\tabularnewline
 &  &  & $m_{4}$ &  & 1.0865 & 1.0866 & 1.0891 & 1.0890 & 1.2642 & 1.2642\tabularnewline
 &  &  &  &  &  &  &  &  &  & \tabularnewline
$\alpha=0^{0},$ &  & 0.0095 & $m_{1}$ &  & 0.01814 & 0.01850 & 0.07235 & 0.07201 & 0.58099 & 0.58099\tabularnewline
$\beta=180^{0}$ &  &  & $m_{2}$ &  & 0.02124 & 0.02171 & 0.07300 & 0.07262 & 0.58108 & 0.58108\tabularnewline
$\gamma=180^{0}$ &  &  & $m_{3}$ &  & 0.04974 & 0.05018 & 0.08636 & 0.08560 & 0.58322 & 0.58322\tabularnewline
 &  &  & $m_{4}$ &  & 1.0865 & 1.0865 & 1.0890 & 1.0888 & 1.2595 & 1.2595\tabularnewline
 &  & 0.011 & $m_{1}$ &  & 0.01918 & 0.02001 & 0.07251 & 0.07217 & 0.58301 & 0.58301\tabularnewline
 &  &  & $m_{2}$ &  & 0.02197 & 0.02342 & 0.07318 & 0.07279 & 0.58309 & 0.58309\tabularnewline
 &  &  & $m_{3}$ &  & 0.04997 & 0.05120 & 0.08671 & 0.08594 & 0.58524 & 0.58524\tabularnewline
 &  &  & $m_{4}$ &  & 1.0865 & 1.0865 & 1.0891 & 1.0889 & 1.2617 & 1.2617\tabularnewline
 &  & 0.0125 & $m_{1}$ &  & 0.02194 & 0.02036 & 0.07269 & 0.07235 & 0.58531 & 0.58531\tabularnewline
 &  &  & $m_{2}$ &  & 0.02432 & 0.02374 & 0.07338 & 0.07299 & 0.58540 & 0.58540\tabularnewline
 &  &  & $m_{3}$ &  & 0.05122 & 0.05146 & 0.08712 & 0.08633 & 0.58755 & 0.58755\tabularnewline
 &  &  & $m_{4}$ &  & 1.0865 & 1.0866 & 1.0891 & 1.0890 & 1.2642 & 1.2642\tabularnewline
\hline 
\end{tabular}
\end{table}

\begin{table}
\caption{Neutrino mass eigenvalues in eV for Normal mass order}

\begin{tabular}{lllllllllll}
 &  &  &  &  &  &  &  &  &  & \tabularnewline
\hline 
Majorana  &  & Effective mass & Mass  &  &  &  & Scheme &  &  & \tabularnewline
\cline{6-11} \cline{7-11} \cline{8-11} \cline{9-11} \cline{10-11} \cline{11-11} 
Phases &  & ($m_{e}$ in eV) & states &  & 1 & 2 & 3 & 4 & 5 & 6\tabularnewline
\hline 
 &  &  &  &  &  &  &  &  &  & \tabularnewline
$\alpha=180^{0},$ &  & 0.0095 & $m_{1}$ &  & 0.01642 & 0.01697 & 0.07235 & 0.07201 & 0.58099 & 0.58099\tabularnewline
$\beta=0^{0},$ &  &  & $m_{2}$ &  & 0.01965 & 0.02034 & 0.07300 & 0.07262 & 0.58107 & 0.58107\tabularnewline
$\gamma=0^{0}$ &  &  & $m_{3}$ &  & 0.04887 & 0.04948 & 0.08636 & 0.08560 & 0.58322 & 0.58322\tabularnewline
 &  &  & $m_{4}$ &  & 1.0865 & 1.0865 & 1.0890 & 1.0888 & 1.2595 & 1.2595\tabularnewline
 &  & 0.011 & $m_{1}$ &  & 0.01782 & 0.01833 & 0.07251 & 0.07217 & 0.58601 & 0.58601\tabularnewline
 &  &  & $m_{2}$ &  & 0.02069 & 0.02192 & 0.07318 & 0.07279 & 0.58309 & 0.58309\tabularnewline
 &  &  & $m_{3}$ &  & 0.04934 & 0.05041 & 0.08671 & 0.08594 & 0.58524 & 0.58524\tabularnewline
 &  &  & $m_{4}$ &  & 1.08649 & 1.0865 & 1.0891 & 1.0889 & 1.2617 & 1.2617\tabularnewline
 &  & 0.0125 & $m_{1}$ &  & 0.02050 & 0.02036 & 0.07269 & 0.07235 & 0.58539 & 0.58539\tabularnewline
 &  &  & $m_{2}$ &  & 0.02299 & 0.02374 & 0.07338 & 0.07299 & 0.58540 & 0.58540\tabularnewline
 &  &  & $m_{3}$ &  & 0.05049 & 0.05146 & 0.08712 & 0.08633 & 0.58755 & 0.58755\tabularnewline
 &  &  & $m_{4}$ &  & 1.0865 & 1.0866 & 1.0891 & 1.0890 & 1.2642 & 1.2642\tabularnewline
 &  &  &  &  &  &  &  &  &  & \tabularnewline
$\alpha=180^{0},$ &  & 0.0095 & $m_{1}$ &  & 0.01642 & 0.01697 & 0.07235 & 0.07201 & 0.58099 & 0.58099\tabularnewline
$\beta=0^{0},$ &  &  & $m_{2}$ &  & 0.01965 & 0.02034 & 0.07300 & 0.07262 & 0.58107 & 0.58107\tabularnewline
$\gamma=180^{0}$ &  &  & $m_{3}$ &  & 0.04887 & 0.04948 & 0.08636 & 0.08560 & 0.58322 & 0.58322\tabularnewline
 &  &  & $m_{4}$ &  & 1.0865 & 1.0865 & 1.0890 & 1.0888 & 1.2595 & 1.2595\tabularnewline
 &  & 0.011 & $m_{1}$ &  & 0.01918 & 0.01833 & 0.07251 & 0.07217 & 0.58601 & 0.58601\tabularnewline
 &  &  & $m_{2}$ &  & 0.02193 & 0.02192 & 0.07318 & 0.07279 & 0.58301 & 0.58301\tabularnewline
 &  &  & $m_{3}$ &  & 0.04997 & 0.05041 & 0.08671 & 0.08594 & 0.58524 & 0.58524\tabularnewline
 &  &  & $m_{4}$ &  & 1.0865 & 1.0865 & 1.0891 & 1.0889 & 1.2617 & 1.2617\tabularnewline
 &  & 0.0125 & $m_{1}$ &  & 0.02050 & 0.02036 & 0.07269 & 0.07235 & 0.58539 & 0.58539\tabularnewline
 &  &  & $m_{2}$ &  & 0.02299 & 0.02374 & 0.07338 & 0.07299 & 0.58540 & 0.58540\tabularnewline
 &  &  & $m_{3}$ &  & 0.05049 & 0.05146 & 0.08712 & 0.08633 & 0.58755 & 0.58755\tabularnewline
 &  &  & $m_{4}$ &  & 1.0865 & 1.0866 & 1.0891 & 1.0890 & 1.2642 & 1.2642\tabularnewline
 &  &  &  &  &  &  &  &  &  & \tabularnewline
$\alpha=180^{0}$ &  & 0.0095 & $m_{1}$ &  & 0.01814 & 0.01850 & 0.07235 & 0.07201 & 0.30781 & 0.30781\tabularnewline
$\beta=180^{0}$ &  &  & $m_{2}$ &  & 0.02124 & 0.02171 & 0.07300 & 0.07262 & 0.30794 & 0.30794\tabularnewline
$\gamma=0^{0}$ &  &  & $m_{3}$ &  & 0.04974 & 0.05018 & 0.08636 & 0.08560 & 0.31104 & 0.31104\tabularnewline
 &  &  & $m_{4}$ &  & 1.0865 & 1.0865 & 1.0890 & 1.0888 & 1.12905 & 1.2905\tabularnewline
 &  & 0.011 & $m_{1}$ &  & 0.02100 & 0.02001 & 0.07251 & 0.07217 & 0.30759 & 0.30759\tabularnewline
 &  &  & $m_{2}$ &  & 0.02363 & 0.02342 & 0.07318 & 0.07279 & 0.30772 & 0.30772\tabularnewline
 &  &  & $m_{3}$ &  & 0.05097 & 0.05120 & 0.08671 & 0.08594 & 0.31082 & 0.31082\tabularnewline
 &  &  & $m_{4}$ &  & 1.0866 & 1.0865 & 1.0891 & 1.0889 & 1.2899 & 1.2899\tabularnewline
 &  & 0.0125 & $m_{1}$ &  & 0.02194 & 0.02214 & 0.07269 & 0.07235 & 0.30734 & 0.30734\tabularnewline
 &  &  & $m_{2}$ &  & 0.02432 & 0.02535 & 0.07338 & 0.07299 & 0.30747 & 0.30747\tabularnewline
 &  &  & $m_{3}$ &  & 0.05122 & 0.05236 & 0.08712 & 0.08633 & 0.31058 & 0.31058\tabularnewline
 &  &  & $m_{4}$ &  & 1.0865 & 1.0866 & 1.0891 & 1.0890 & 1.1289 & 1.1289\tabularnewline
 &  &  &  &  &  &  &  &  &  & \tabularnewline
$\alpha=180^{0}$ &  & 0.0095 & $m_{1}$ &  & 0.01814 & 0.01850 & 0.07235 & 0.07201 & 0.30781 & 0.30781\tabularnewline
$\beta=180^{0}$ &  &  & $m_{2}$ &  & 0.02124 & 0.02171 & 0.07300 & 0.07262 & 0.30794 & 0.30794\tabularnewline
$\gamma=180^{0}$ &  &  & $m_{3}$ &  & 0.04974 & 0.05018 & 0.08636 & 0.08560 & 0.31104 & 0.31104\tabularnewline
 &  &  & $m_{4}$ &  & 1.0865 & 1.0865 & 1.0890 & 1.0888 & 1.1290 & 1.2905\tabularnewline
 &  & 0.011 & $m_{1}$ &  & 0.01918 & 0.02001 & 0.07251 & 0.07217 & 0.30759 & 0.30759\tabularnewline
 &  &  & $m_{2}$ &  & 0.02193 & 0.02342 & 0.07318 & 0.07279 & 0.30772 & 0.30772\tabularnewline
 &  &  & $m_{3}$ &  & 0.04997 & 0.05120 & 0.08671 & 0.08594 & 0.31082 & 0.31082\tabularnewline
 &  &  & $m_{4}$ &  & 1.0866 & 1.0865 & 1.0891 & 1.0889 & 1.2899 & 1.2899\tabularnewline
 &  & 0.0125 & $m_{1}$ &  & 0.02194 & 0.02214 & 0.07269 & 0.07235 & 0.30734 & 0.30734\tabularnewline
 &  &  & $m_{2}$ &  & 0.02432 & 0.02535 & 0.07338 & 0.07299 & 0.30747 & 0.30747\tabularnewline
 &  &  & $m_{3}$ &  & 0.05122 & 0.05236 & 0.08712 & 0.08633 & 0.31058 & 0.31058\tabularnewline
 &  &  & $m_{4}$ &  & 1.0865 & 1.0866 & 1.0891 & 1.0890 & 1.1289 & 1.1289\tabularnewline
\hline 
\end{tabular}
\end{table}

\begin{table}
\caption{Neutrino mass eigenvalues in eV for Inverted mass order}

\begin{tabular}{lllllllllll}
 &  &  &  &  &  &  &  &  &  & \tabularnewline
\hline 
Majorana  &  & Effective mass & Mass  &  &  &  & Scheme &  &  & \tabularnewline
\cline{6-11} \cline{7-11} \cline{8-11} \cline{9-11} \cline{10-11} \cline{11-11} 
Phases &  & ($m_{e}$ in eV) & states &  & 1 & 2 & 3 & 4 & 5 & 6\tabularnewline
\hline 
 &  &  &  &  &  &  &  &  &  & \tabularnewline
$\alpha=0^{0},$ &  & 0.03 & $m_{1}$ &  & 0.05242 & 0.04765 & 0.31268 & 0.31187 & 0.67474 & 0.67474\tabularnewline
$\beta=0^{0}$ &  &  & $m_{2}$ &  & 0.04501 & 0.03934 & 0.31153 & 0.31072 & 0.67421 & 0.67421\tabularnewline
$\gamma=0^{0}$ &  &  & $m_{3}$ &  & 0.02703 & 0.01436 & 0.30947 & 0.30865 & 0.67326 & 0.67326\tabularnewline
 &  &  & $m_{4}$ &  & 1.0866 & 1.0864 & 1.1295 & 1.1293 & 1.2780 & 1.2780\tabularnewline
 &  & 0.04 & $m_{1}$ &  & 0.04863 & 0.04999 & 0.31196 & 0.30907 & 0.67510 & 0.67510\tabularnewline
 &  &  & $m_{2}$ &  & 0.04056 & 0.04211 & 0.31080 & 0.30790 & 0.67451 & 0.67451\tabularnewline
 &  &  & $m_{3}$ &  & 0.01908 & 0.02082 & 0.30873 & 0.30582 & 0.67346 & 0.67346\tabularnewline
 &  &  & $m_{4}$ &  & 1.0864 & 1.0865 & 1.1293 & 1.1285 & 1.291 & 1.291\tabularnewline
 &  & 0.05 & $m_{1}$ &  & 0.05406 & 0.05886 & 0.31160 & 0.30753 & 0.37923 & 0.37923\tabularnewline
 &  &  & $m_{2}$ &  & 0.04686 & 0.05234 & 0.31044 & 0.30636 & 0.37800 & 0.37800\tabularnewline
 &  &  & $m_{3}$ &  & 0.02965 & 0.03796 & 0.30837 & 0.30426 & 0.37580 & 0.37580\tabularnewline
 &  &  & $m_{4}$ &  & 1.0867 & 1.0870 & 1.1292 & 1.1281 & 1.1662 & 1.1662\tabularnewline
 &  &  &  &  &  &  &  &  &  & \tabularnewline
$\alpha=0^{0},$ &  & 0.03 & $m_{1}$ &  & 0.05242 & 0.04765 & 0.31268 & 0.31187 & 0.67474 & 0.67474\tabularnewline
$\beta=0^{0}$ &  &  & $m_{2}$ &  & 0.04501 & 0.03934 & 0.31153 & 0.31072 & 0.67421 & 0.67421\tabularnewline
$\gamma=180^{0}$ &  &  & $m_{3}$ &  & 0.02703 & 0.01436 & 0.30947 & 0.30865 & 0.67326 & 0.67326\tabularnewline
 &  &  & $m_{4}$ &  & 1.0866 & 1.0864 & 1.1295 & 1.1293 & 1.2780 & 1.2780\tabularnewline
 &  & 0.04 & $m_{1}$ &  & 0.04863 & 0.04999 & 0.31196 & 0.30907 & 0.67510 & 0.67510\tabularnewline
 &  &  & $m_{2}$ &  & 0.04056 & 0.04211 & 0.31080 & 0.30790 & 0.67451 & 0.67451\tabularnewline
 &  &  & $m_{3}$ &  & 0.01908 & 0.02082 & 0.30873 & 0.30582 & 0.67346 & 0.67346\tabularnewline
 &  &  & $m_{4}$ &  & 1.0864 & 1.0865 & 1.1293 & 1.1285 & 1.228 & 1.228\tabularnewline
 &  & 0.05 & $m_{1}$ &  & 0.05406 & 0.05886 & 0.31160 & 0.30753 & 0.37923 & 0.37923\tabularnewline
 &  &  & $m_{2}$ &  & 0.04686 & 0.05234 & 0.31044 & 0.30636 & 0.37800 & 0.37800\tabularnewline
 &  &  & $m_{3}$ &  & 0.02965 & 0.03796 & 0.30837 & 0.30426 & 0.37580 & 0.37580\tabularnewline
 &  &  & $m_{4}$ &  & 1.0867 & 1.0870 & 1.1292 & 1.1281 & 1.285 & 1.285\tabularnewline
 &  &  &  &  &  &  &  &  &  & \tabularnewline
$\alpha=0^{0},$ &  & 0.03 & $m_{1}$ &  & 0.05242 & 0.04765 & 0.31268 & 0.31187 & 0.67474 & 0.67474\tabularnewline
$\beta=180^{0}$ &  &  & $m_{2}$ &  & 0.04501 & 0.03934 & 0.31153 & 0.31072 & 0.67421 & 0.67421\tabularnewline
$\gamma=0^{0}$ &  &  & $m_{3}$ &  & 0.02703 & 0.01436 & 0.30947 & 0.30865 & 0.67326 & 0.67326\tabularnewline
 &  &  & $m_{4}$ &  & 1.0866 & 1.0864 & 1.1295 & 1.1293 & 1.2780 & 1.2780\tabularnewline
 &  & 0.04 & $m_{1}$ &  & 0.04863 & 0.04999 & 0.31196 & 0.30907 & 0.67510 & 0.67510\tabularnewline
 &  &  & $m_{2}$ &  & 0.04056 & 0.04211 & 0.31080 & 0.30790 & 0.67451 & 0.67451\tabularnewline
 &  &  & $m_{3}$ &  & 0.01908 & 0.02082 & 0.30873 & 0.30582 & 0.67346 & 0.67346\tabularnewline
 &  &  & $m_{4}$ &  & 1.0864 & 1.0865 & 1.1293 & 1.1285 & 1.2906 & 1.2906\tabularnewline
 &  & 0.05 & $m_{1}$ &  & 0.05406 & 0.05886 & 0.31160 & 0.30754 & 0.37923 & 0.37923\tabularnewline
 &  &  & $m_{2}$ &  & 0.04686 & 0.05234 & 0.31044 & 0.30637 & 0.37800 & 0.37800\tabularnewline
 &  &  & $m_{3}$ &  & 0.02965 & 0.03796 & 0.30837 & 0.30427 & 0.37580 & 0.37580\tabularnewline
 &  &  & $m_{4}$ &  & 1.0867 & 1.0870 & 1.1292 & 1.1281 & 1.1662 & 1.1662\tabularnewline
 &  &  &  &  &  &  &  &  &  & \tabularnewline
$\alpha=0^{0},$ &  & 0.03 & $m_{1}$ &  & 0.05242 & 0.04765 & 0.31268 & 0.31187 & 0.67474 & 0.67474\tabularnewline
$\beta=180^{0}$ &  &  & $m_{2}$ &  & 0.04501 & 0.03934 & 0.31153 & 0.31072 & 0.67421 & 0.67421\tabularnewline
$\gamma=180^{0}$ &  &  & $m_{3}$ &  & 0.02703 & 0.01436 & 0.30947 & 0.30865 & 0.67326 & 0.67326\tabularnewline
 &  &  & $m_{4}$ &  & 1.0866 & 1.0864 & 1.1295 & 1.1293 & 1.2780 & 1.2780\tabularnewline
 &  & 0.04 & $m_{1}$ &  & 0.04863 & 0.04999 & 0.31196 & 0.30907 & 0.67510 & 0.67510\tabularnewline
 &  &  & $m_{2}$ &  & 0.04056 & 0.04211 & 0.31080 & 0.30790 & 0.67451 & 0.67451\tabularnewline
 &  &  & $m_{3}$ &  & 0.01908 & 0.02082 & 0.30873 & 0.30582 & 0.67346 & 0.67346\tabularnewline
 &  &  & $m_{4}$ &  & 1.0864 & 1.0865 & 1.1293 & 1.1285 & 1.2282 & 1.2282\tabularnewline
 &  & 0.05 & $m_{1}$ &  & 0.05406 & 0.05886 & 0.31160 & 0.30754 & 0.37923 & 0.37923\tabularnewline
 &  &  & $m_{2}$ &  & 0.04686 & 0.05234 & 0.31044 & 0.30637 & 0.37800 & 0.37800\tabularnewline
 &  &  & $m_{3}$ &  & 0.02965 & 0.03796 & 0.30837 & 0.30427 & 0.37580 & 0.37580\tabularnewline
 &  &  & $m_{4}$ &  & 1.0867 & 1.0870 & 1.1292 & 1.1281 & 1.2849 & 1.2849\tabularnewline
\hline 
\end{tabular}
\end{table}

\begin{table}
\caption{Neutrino mass eigenvalues in eV for Inverted mass order}

\begin{tabular}{lllllllllll}
 &  &  &  &  &  &  &  &  &  & \tabularnewline
\hline 
Majorana  &  & Effective mass & Mass  &  &  &  & Scheme &  &  & \tabularnewline
\cline{6-11} \cline{7-11} \cline{8-11} \cline{9-11} \cline{10-11} \cline{11-11} 
Phases &  & ($m_{e}$ in eV) & states &  & 1 & 2 & 3 & 4 & 5 & 6\tabularnewline
\hline 
 &  &  &  &  &  &  &  &  &  & \tabularnewline
$\alpha=180^{0},$ &  & 0.03 & $m_{1}$ &  & 0.06594 & 0.06814 & 0.29638 & 0.30574 & 0.13709 & 0.13709\tabularnewline
$\beta=0^{0},$ &  &  & $m_{2}$ &  & 0.05981 & 0.06186 & 0.29516 & 0.30455 & 0.13408 & 0.13408\tabularnewline
$\gamma=0^{0}$ &  &  & $m_{3}$ &  & 0.04599 & 0.04526 & 0.29298 & 0.30245 & 0.12844 & 0.12844\tabularnewline
 &  &  & $m_{4}$ &  & 1.0875 & 1.0877 & 1.1252 & 1.1276 & 1.0949 & 1.0949\tabularnewline
 &  & 0.04 & $m_{1}$ &  & 0.07885 & 0.08259 & 0.28885 & 0.30241 & 0.13859 & 0.13859\tabularnewline
 &  &  & $m_{2}$ &  & 0.07278 & 0.07705 & 0.28759 & 0.30122 & 0.13530 & 0.13530\tabularnewline
 &  &  & $m_{3}$ &  & 0.05882 & 0.06341 & 0.28535 & 0.29909 & 0.12897 & 0.12897\tabularnewline
 &  &  & $m_{4}$ &  & 1.0886 & 1.0890 & 1.1233 & 1.1276 & 1.0958 & 1.0958\tabularnewline
 &  & 0.05 & $m_{1}$ &  & 0.12418 & 0.13887 & 0.26689 & 0.27993 & 0.14073 & 0.14073\tabularnewline
 &  &  & $m_{2}$ &  & 0.12097 & 0.13622 & 0.26554 & 0.27864 & 0.13702 & 0.13702\tabularnewline
 &  &  & $m_{3}$ &  & 0.11480 & 0.13138 & 0.26310 & 0.27633 & 0.12954 & 0.12954\tabularnewline
 &  &  & $m_{4}$ &  & 1.0927 & 1.0943 & 1.1178 & 1.1209 & 1.0969 & 1.0969\tabularnewline
 &  &  &  &  &  &  &  &  &  & \tabularnewline
$\alpha=180^{0},$ &  & 0.03 & $m_{1}$ &  & 0.06594 & 0.06814 & 0.29638 & 0.30574 & 0.13709 & 0.13709\tabularnewline
$\beta=0^{0},$ &  &  & $m_{2}$ &  & 0.05981 & 0.06186 & 0.29516 & 0.30455 & 0.13408 & 0.13408\tabularnewline
$\gamma=180^{0}$ &  &  & $m_{3}$ &  & 0.04599 & 0.04526 & 0.29298 & 0.30245 & 0.12844 & 0.12844\tabularnewline
 &  &  & $m_{4}$ &  & 1.0875 & 1.0877 & 1.1252 & 1.1276 & 1.0949 & 1.0949\tabularnewline
 &  & 0.04 & $m_{1}$ &  & 0.07885 & 0.08259 & 0.28885 & 0.30241 & 0.13859 & 0.13859\tabularnewline
 &  &  & $m_{2}$ &  & 0.07278 & 0.07705 & 0.28759 & 0.30122 & 0.13530 & 0.13530\tabularnewline
 &  &  & $m_{3}$ &  & 0.05882 & 0.06341 & 0.28535 & 0.29909 & 0.12897 & 0.12897\tabularnewline
 &  &  & $m_{4}$ &  & 1.0886 & 1.0890 & 1.1233 & 1.1259 & 1.0958 & 1.0958\tabularnewline
 &  & 0.05 & $m_{1}$ &  & 0.13689 & 0.14226 & 0.27806 & 0.27803 & 0.14073 & 0.14073\tabularnewline
 &  &  & $m_{2}$ &  & 0.13419 & 0.13968 & 0.27675 & 0.27673 & 0.13702 & 0.13702\tabularnewline
 &  &  & $m_{3}$ &  & 0.12926 & 0.13495 & 0.27441 & 0.27441 & 0.12954 & 0.12954\tabularnewline
 &  &  & $m_{4}$ &  & 1.0949 & 1.0948 & 1.1207 & 1.1204 & 1.0969 & 1.0969\tabularnewline
 &  &  &  &  &  &  &  &  &  & \tabularnewline
$\alpha=180^{0}$ &  & 0.03 & $m_{1}$ &  & 0.06594 & 0.06814 & 0.29638 & 0.30574 & 0.13709 & 0.13709\tabularnewline
$\beta=180^{0}$ &  &  & $m_{2}$ &  & 0.05981 & 0.06186 & 0.29516 & 0.30455 & 0.13408 & 0.13408\tabularnewline
$\gamma=0^{0}$ &  &  & $m_{3}$ &  & 0.04599 & 0.04526 & 0.29298 & 0.30245 & 0.12844 & 0.12844\tabularnewline
 &  &  & $m_{4}$ &  & 1.0875 & 1.0877 & 1.1252 & 1.1276 & 1.0949 & 1.0949\tabularnewline
 &  & 0.04 & $m_{1}$ &  & 0.07885 & 0.08259 & 0.28885 & 0.30241 & 0.13859 & 0.13859\tabularnewline
 &  &  & $m_{2}$ &  & 0.07278 & 0.07705 & 0.28759 & 0.30122 & 0.13530 & 0.13530\tabularnewline
 &  &  & $m_{3}$ &  & 0.05882 & 0.06341 & 0.28535 & 0.29909 & 0.12897 & 0.12897\tabularnewline
 &  &  & $m_{4}$ &  & 1.0886 & 1.0890 & 1.1233 & 1.1259 & 1.0958 & 1.0958\tabularnewline
 &  & 0.05 & $m_{1}$ &  & 0.12418 & 0.13887 & 0.26689 & 0.27993 & 0.14073 & 0.14073\tabularnewline
 &  &  & $m_{2}$ &  & 0.12097 & 0.13622 & 0.26553 & 0.27864 & 0.13702 & 0.13702\tabularnewline
 &  &  & $m_{3}$ &  & 0.11480 & 0.13138 & 0.26310 & 0.27633 & 0.12954 & 0.12954\tabularnewline
 &  &  & $m_{4}$ &  & 1.0927 & 1.0943 & 1.1178 & 1.1208 & 1.0969 & 1.0969\tabularnewline
 &  &  &  &  &  &  &  &  &  & \tabularnewline
$\alpha=180^{0}$ &  & 0.03 & $m_{1}$ &  & 0.06594 & 0.06814 & 0.29638 & 0.30574 & 0.13709 & 0.13709\tabularnewline
$\beta=180^{0}$ &  &  & $m_{2}$ &  & 0.05981 & 0.06186 & 0.29516 & 0.30455 & 0.13408 & 0.13408\tabularnewline
$\gamma=180^{0}$ &  &  & $m_{3}$ &  & 0.04599 & 0.04526 & 0.29298 & 0.30245 & 0.12844 & 0.12844\tabularnewline
 &  &  & $m_{4}$ &  & 1.0875 & 1.0877 & 1.1252 & 1.1276 & 1.0949 & 1.0949\tabularnewline
 &  & 0.04 & $m_{1}$ &  & 0.07885 & 0.08259 & 0.28885 & 0.30241 & 0.13859 & 0.13859\tabularnewline
 &  &  & $m_{2}$ &  & 0.07278 & 0.07705 & 0.28759 & 0.30122 & 0.13530 & 0.13530\tabularnewline
 &  &  & $m_{3}$ &  & 0.05882 & 0.06341 & 0.28535 & 0.29909 & 0.12897 & 0.12897\tabularnewline
 &  &  & $m_{4}$ &  & 1.0886 & 1.0890 & 1.1233 & 1.1259 & 1.0958 & 1.0958\tabularnewline
 &  & 0.05 & $m_{1}$ &  & 0.13689 & 0.14226 & 0.27806 & 0.27803 & 0.15975 & 0.15975\tabularnewline
 &  &  & $m_{2}$ &  & 0.13419 & 0.13968 & 0.27675 & 0.27673 & 0.15648 & 0.15648\tabularnewline
 &  &  & $m_{3}$ &  & 0.12926 & 0.13495 & 0.27441 & 0.27441 & 0.14994 & 0.14994\tabularnewline
 &  &  & $m_{4}$ &  & 1.0949 & 1.0948 & 1.1207 & 1.1204 & 1.0997 & 1.0997\tabularnewline
\hline 
\end{tabular}
\end{table}

\section{Conclusions }

In the future, the aim of the neutrinoless double beta decay experiment
would predict the constraint on the limit of the lightest neutrino
mass ~$\,0.0007\,eV\leq m_{1}\leq0.0008\,eV\,$ and mass order by
reaching the high sensitivity of 0.001 eV. We have study the neutrino
mass eigenvalue in the case of both neutrino mass order by using six
different possible scheme. In table (1) and (2), by taking the effective
neutrino mass,$\,m_{e}=(0.0095,\,0.011,\,0.0125)\,eV,\,$within normal
neutrino mass order {[}29{]} and for scheme (1-4), we have found that
$\,m_{1}\,$varied from (0.016--0.072) eV,$\,m_{2}\,$varied from
(0.019--0.073) eV, $\,m_{3}\,$varied from (0.048-0.086) eV and $\,m_{4}\,$varied
from (1.08-1.09) eV for different allowed range of majorana phase.
Since equations for scheme 5 and 6 for normal order (see eq. (24-25))
are almost nearly the same and the only difference lies in degenerate
masses $\,\Delta m_{atm}^{2}\,and\,\Delta m_{solar\,}^{2},$which
provide the subtle effects into equations {[}30, 31, 32{]}. Thus we
have found nearly same mass eigenvalues for both schemes i.e $\,m_{(2+2)}\equiv\left(0.30-1.13\right)eV$.
We also find limit of the sum of all four neutrino masses $\text{\,}\text{\ensuremath{\sum_{(3+1)}\equiv}}m_{1}+m_{2}+m_{3}+m_{4}=(1.17-1.32)\,eV\,$and~$\text{\,}\sum_{(2+2)}\equiv m_{1}+m_{2}+m_{3}+m_{4}=(2.0-3.0)\,eV\,$~relevant
for cosmological observations. In the case of inverted neutrino mass
order {[}30{]}, by assuming the value of the effective neutrino mass~$m_{e}=(0.03-0.05)\,eV,$
we have found that $\,m_{1}\,$varied from (0.045--0.278) eV,$\,$varied
from (0.040--0.276) eV, $\,m_{3}\,$varied from (0.019-0.274) eV
and $\,m_{4}\,$varied from (1.08--1.12) eV as given in table (3)
and (4) and nearly same mass eigenvalues for schemes 5 and 6 i.e $m_{(2+2)}\equiv\left(0.137-1.28\right)eV$~due
to the subtle effect produce into the equations by the degenerate
masses $\,\Delta m_{atm}^{2}\,and\,\Delta m_{solar\,}^{2},\,$for
different allowed range of majorana phase. We find limit of the sum
of all four neutrino masses $\sum_{(3+1)}\text{\ensuremath{\equiv}}m_{1}+m_{2}+m_{3}+m_{4}=(1.18-1.23)\,eV\,$and~$\text{\,}\sum_{(2+2)}\equiv m_{1}+m_{2}+m_{3}+m_{4}=(1.49-3.33)\,eV\,$~
relevant for cosmological observations. The value of effective Majorana
mass $\,|m_{e}|\,$ as a function of lower mass $\,m_{1}\,$and sum
of the neutrino masses $\,\sum_{(3+1)}\,$ in the case of (3+1) neutrino
mixing with Normal and Inverted Ordering are shown in Left and right
pannel of fig. (3), respectively. The inverted mass order provides
the smaller window of sum of all four masses for (2+2) scheme than
the (3+1)scheme with larger value of mass. Indeed, for the (3+1) scheme
the mass of sterile neutrino is just simply add up to the former active
3 neutrinos while (2+2) scheme the sterile neutrino is just distroyed
the whole mass spectrum. The neutrino mass eigen state is depend on
choice of effective neutrino mass and allowed range of majorana phase.
Hence precise determination of effective neutrino mass from beta decay
experiment will gives the exact picture of mass spectrum. 

\begin{figure}
\includegraphics[width=9.5cm,height=10cm]{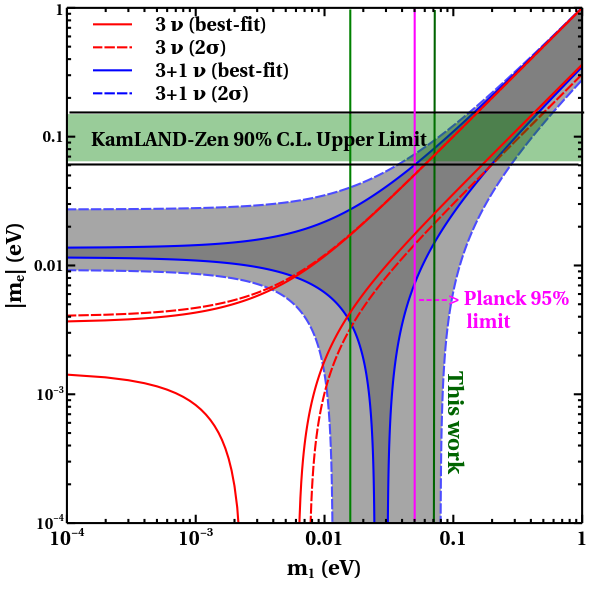} \includegraphics[width=9.5cm,height=10cm]{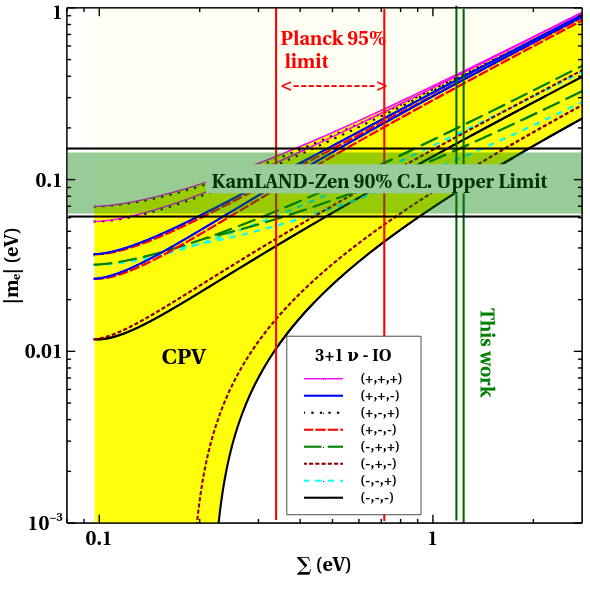}

\caption{Left and right pannel respectively shows the value of effective Majorana
mass $\,|m_{e}|\,$ as a function of lower mass $\,m_{1}\,$and sum
of the neutrino masses $\,\sum_{(3+1)}\,$ in the case of (3+1) neutrino
mixing with Normal and Inverted Ordering. The signs combination as
shown in the legend imply the signs of $\,e^{i\alpha},\,e^{i\beta},\,e^{i\gamma}=\pm1\,$
for the eight possible cases in which CP is conserved.}
\end{figure}

\end{document}